\documentclass[aps,pra,preprint,showpacs]{revtex4}

\usepackage[dvips]{graphicx}

\begin{document}

\title{A new perturbation method in quantum mechanics}

\author{Francisco M. Fern\'{a}ndez}

\affiliation{INIFTA (Conicet, UNLP), Diag 113 y 64, S/N, Sucursal
4, Casilla de Correo 16, 1900 La Plata, Argentina}

\email{fernande@quimica.unlp.edu.ar}

\begin{abstract}
We investigate the convergence properties of a perturbation method
proposed some time ago and reveal some of it most interesting
features. Anharmonic oscillators in the strong--coupling limit
prove to be appropriate illustrative examples and benchmark.
\end{abstract}


\maketitle

\section{\label{sec:Intro}Introduction}

Some time ago Bessis and Bessis (BB from now on) \cite{BB97} proposed a new
perturbation approach in quantum mechanics based on the application of a
factorization method to a Riccati equation derived from the Schr\"{o}dinger
one. They obtained reasonable results for some of the energies of the
quartic anharmonic oscillator by means of perturbation series of order
fourth and sixth, without resorting to a resummation method. In spite of
this success, BB's method has passed unnoticed as far as we know.

The purpose of this paper is to investigate BB's perturbation method in more
detail. In Section \ref{sec:Method} we write it in a quite general way and
derive other approaches as particular cases. In Section \ref{sec:Results} we
carry out perturbation calculations of sufficiently large order and try to
find out numerical evidence of convergence. One--dimensional anharmonic
oscillators prove to be a suitable benchmark for present numerical tests.
For simplicity we restrict to ground states and  choose straightforward
logarithmic perturbation theory instead of the factorization method proposed
by BB \cite{BB97}. Finally, in Section \ref{sec:Conclusions} we discuss the
results and draw some conclusions.

\section{\label{sec:Method}The method}

In standard Rayleigh--Schr\"{o}dinger perturbation theory we try to solve
the eigenvalue equation
\begin{equation}
\hat{H}\Psi =E\Psi ,\;\hat{H}=\hat{H}_{0}+\lambda \hat{H}^{\prime }
\label{eq:Schrodinger}
\end{equation}
by expanding the energy $E$ and eigenfunction $\Psi $ in a Taylor series
about $\lambda =0$. This method is practical provided that we can solve the
eigenvalue equation for $\lambda =0$.

In some cases it is more convenient to construct a parameter--dependent
Hamiltonian operator $\hat{H}(\beta )$ that one can expand in a Taylor
series about $\beta =0$
\begin{equation}
\hat{H}(\beta )=\sum_{j=0}\hat{H}_{j}\beta ^{j}  \label{eq:H(beta)_series}
\end{equation}
in such a way that we can solve the eigenvalue equation for $\hat{H}(0)=\hat{%
H}_{0}$. In this case we expand the eigenfunctions $\Psi (\beta )$ and
eigenvalues $E(\beta )$ in Taylor series:
\begin{equation}
\Psi (\beta )=\sum_{j=0}^{\infty }\Psi _{j}\beta ^{j},\;E(\beta
)=\sum_{j=0}^{\infty }E_{j}\beta ^{j}  \label{eq:Psi,E_series}
\end{equation}
There are many practical examples of application of this alternative
approach \cite{F01}. In particular, BB \cite{BB97} suggested the following
form of $\hat{H}(\beta )$:
\begin{eqnarray}
\hat{H}(\beta ) &=&\hat{H}+(\beta -1)\hat{W}(\beta )  \nonumber \\
\hat{W}(\beta ) &=&\sum_{j=0}\hat{W}_{j}\beta ^{j}.
\label{eq:H(beta)_Bessis}
\end{eqnarray}
Comparing equations (\ref{eq:H(beta)_series}) and (\ref{eq:H(beta)_Bessis})
we conclude that
\begin{eqnarray}
\hat{H}_{0} &=&\hat{H}-\hat{W}_{0},  \nonumber \\
\hat{H}_{j} &=&\hat{W}_{j-1}-\hat{W}_{j},\;j>0.  \label{eq:H_j_Bessis}
\end{eqnarray}
In principle there is enormous flexibility in the choice of the operator
coefficients $\hat{W}_{j}$ as we show below by derivation of two known
particular cases.

If we restrict the expansion (\ref{eq:H(beta)_Bessis}) to just one term $%
\hat{W}(\beta )=\hat{W}(0)=\hat{W}_{0}$ then $\hat{H}(\beta )=\hat{H}-\hat{W}%
_{0}+\beta \hat{W}_{0}$. Choosing $\hat{W}_{0}=\hat{H}-\hat{H}_{0}(\alpha )=%
\hat{H}^{\prime }(\alpha )$, where $\hat{H}_{0}(\alpha )$ is a
parameter--dependent Hamiltonian operator with known eigenvalues and
eigenfunctions, we obtain the method proposed by Killingbeck \cite{K81} some
time ago. The main strategy behind this approach is to choose an appropriate
value of the adjustable parameter $\alpha $ leading to a renormalized
perturbation series with the best convergence properties \cite{F01,K81}.

If we consider two terms of the form $\hat{W}_{0}=\hat{H}-\hat{H}_{0}(\alpha
)$ and $\hat{W}_{1}=\lambda \hat{H}^{\prime }$, then we derive the
Hamiltonian operator $\hat{H}(\beta )=\hat{H}_{0}(\alpha )+\beta [\hat{H}%
_{0}-\hat{H}_{0}(\alpha )]+\beta ^{2}\lambda \hat{H}^{\prime }$ that
Killingbeck et al \cite{KGJ01} have recently found to be even more
convenient for the treatment of some perturbation problems. Those approaches
are practical if we can solve the eigenvalue equation for $\beta =0$ and all
relevant values of $\alpha $.

We should mention that it was not the aim of BB to obtain renormalized
series with an adjustable parameter but to choose the operator coefficients $%
\hat{W}_{j}$ in such a way that they could solve the perturbation equations
\begin{equation}
\left( \hat{H}_{0}-E_{0}\right) \Psi _{j}=\sum_{i=1}^{j}\left( E_{i}-\hat{W}%
_{i}+\hat{W}_{i-1}\right) \Psi _{j-i}  \label{eq:PT_j}
\end{equation}
in exact algebraic form \cite{BB97}.

For simplicity in this paper we concentrate on a one--dimensional eigenvalue
equation of the form
\begin{equation}
\Psi ^{^{\prime \prime }}(x)=[U(x)-E]\Psi (x),\;U(x)=V(x)+\frac{l(l+1)}{x^{2}%
}.  \label{eq:Schro_x}
\end{equation}
If $\Psi (0)=\Psi (\infty )=0$ and $l=0,1,2,\ldots $ is the
angular--momentum quantum number, this equation applies to central--field
models. If $\Psi (-\infty )=\Psi (\infty )=0$ and $l=-1$, we have a
one--dimensional model. In particular, when $V(x)=V(-x)$ then $l=-1$, and $%
l=0$, select the spaces of even and odd solutions, respectively. In any case
the regular solution to the eigenvalue equation (\ref{eq:Schro_x}) behaves
asymptotically as $x^{l+1}$ at origin.

In order to calculate perturbation corrections of sufficiently large order
by means of BB's method we define
\begin{equation}
f(x)=\frac{s}{x}-\frac{\Psi ^{\prime }(x)}{\Psi (x)},\;s=l+1  \label{eq:f(x)}
\end{equation}
that satisfies the Riccati equation
\begin{equation}
f^{\prime }+\frac{2s}{x}f-f^{2}+V-E=0.  \label{eq:Riccati}
\end{equation}
The corresponding equation for the Hamiltonian $\hat{H}(\beta )$ in equation
(\ref{eq:H(beta)_Bessis}) reads
\begin{equation}
f^{\prime }+\frac{2s}{x}f-f^{2}+V-E+(\beta -1)W=0,  \label{eq:Riccati_beta}
\end{equation}
if we restrict to the case that $\hat{W}(\beta )=W(\beta ,x)$ depends only
on $\beta $ and the coordinate. The coefficients of the expansion
\begin{equation}
f=\sum_{j=0}^{\infty }f_{j}\beta ^{j}  \label{eq:f_series}
\end{equation}
satisfy the perturbation equations
\begin{equation}
f_{j}^{\prime }+\frac{2s}{x}f_{j}-\sum_{k=0}^{j}f_{k}f_{j-k}+V\delta
_{j0}-E_{j}+W_{j-1}-W_{j}=0.  \label{eq:PT_f_j}
\end{equation}

\section{\label{sec:Results}Results}

Simple one--dimensional anharmonic oscillators $V(x)=x^{2}+\lambda x^{2K}$, $%
K=2,3,\ldots $, are a suitable demanding benchmark for testing new
perturbation approaches. We easily increase the degree of difficulty by
increasing the values of the coupling parameter $\lambda $ and anharmonicity
exponent $K$. BB applied their method to the first four energy levels of the
model with $K=2$ and several values of $\lambda $, restricting their
calculation to perturbation theory of order four and six. Here we consider
the strong--coupling limit ($\lambda \rightarrow \infty $) of the
oscillators mentioned above:
\begin{equation}
V(x)=x^{2K}.  \label{eq:x^2K}
\end{equation}
Notice that if the perturbation series gives acceptable results for this
case, then it will certainly be suitable for all $0<\lambda <\infty $.
Moreover, the perturbation corrections for these models are simpler enabling
us to proceed to higher orders with less computational requirement.

In order to make present discussion clearer we first illustrate the main
ideas of the method with the pure quartic oscillator $K=2$. We try
polynomial solutions of the form
\begin{equation}
f_{j}(x)=\sum_{m=0}^{j+1}c_{j,2m+1}x^{2m+1},\;j=0,1,\ldots
\label{eq:f_j_x4}
\end{equation}
in the perturbation equations (\ref{eq:PT_f_j}) for the ground state ($s=0$%
). Substitution of $f_{0}(x)$ into the perturbation equation of order zero
leads to
\begin{equation}
-c_{0,3}^{2}x^{6}+(1-2c_{0,1}c_{0,3})x^{4}+(3c_{0,3}-c_{0,1}^{2})x^{2}+c_{0,1}-E_{0}-W_{0}=0.
\label{eq:PT_0}
\end{equation}
In order to have a solution with $c_{0,3}\neq 0$ we choose $%
W_{0}=-c_{0,3}^{2}x^{6}$; then $c_{0,3}=1/(2c_{0,1})$, and $%
3c_{0,3}-c_{0,1}^{2}=0$ becomes a cubic equation with two complex and one
real root. If we select the later we finally have
\begin{eqnarray}
f_{0} &=&\frac{12^{1/3}}{2}x+12^{-1/3}x^{3},  \nonumber \\
W_{0} &=&-12^{-2/3}x^{6},  \nonumber \\
E_{0} &=&\frac{12^{1/3}}{2}\approx 1.1447.  \label{eq:f0,E0}
\end{eqnarray}
We expect the resulting unperturbed wavefunction
\begin{equation}
\Psi _{0}\propto \exp \left( -\frac{12^{1/3}}{4}x^{2}-\frac{x^{4}}{4\times
12^{1/3}}\right)   \label{eq:Psi_0}
\end{equation}
to be an improvement on the harmonic--oscillator one in standard
Rayleigh--Schr\"{o}dinger perturbation theory \cite{F01,K81,KGJ01}. The
zeroth--order energy (\ref{eq:f0,E0}) is reasonably close to the exact value
shown in Table~\ref{tab:1} which was obtained with the Riccati--Pad\'{e} method
\cite {FMT89}.

At first order we have
\begin{eqnarray}
&-&\frac{12^{2/3}}{6}c_{1,5}x^{8}-12^{1/3}\left( c_{1,5}+\frac{%
12^{1/3}c_{1,3}}{6}+\frac{1}{12}\right) x^{6}  \nonumber \\
&+&\left( 5c_{1,5}-12^{1/3}c_{1,3}-\frac{12^{2/3}c_{1,1}}{6}\right)
x^{4}+\left( 3c_{1,3}-12^{1/3}c_{1,1}\right) x^{2}  \nonumber \\
&+&c_{1,1}-E_{1}-W_{1}=0.  \label{eq:PT_1}
\end{eqnarray}
We easily solve this equation if $W_{1}=-12^{2/3}c_{1,5}x^{8}/6$; the result
is
\begin{eqnarray}
f_{1} &=&-\frac{5}{112}12^{1/3}\left( x+\frac{12^{1/3}}{3}\right) -\frac{%
3x^{5}}{56},  \nonumber \\
W_{1} &=&\frac{12^{2/3}x^{8}}{112},  \nonumber \\
E_{1} &=&-\frac{5}{112}12^{1/3}\approx -0.1022.  \label{eq:f1,E1}
\end{eqnarray}
The energy corrected through first order is somewhat closer to the exact
value: $E_{0}+E_{1}\approx 1.0425$.

\begin{figure}
\caption{$\log|E_j/E_0|$ vs. $j$ for the ground states of the quartic ($K=2$),
sextic ($K=3$ and $K=3$ (b)) and octic ($K=4$) oscillators} \label{fig:1}
\includegraphics{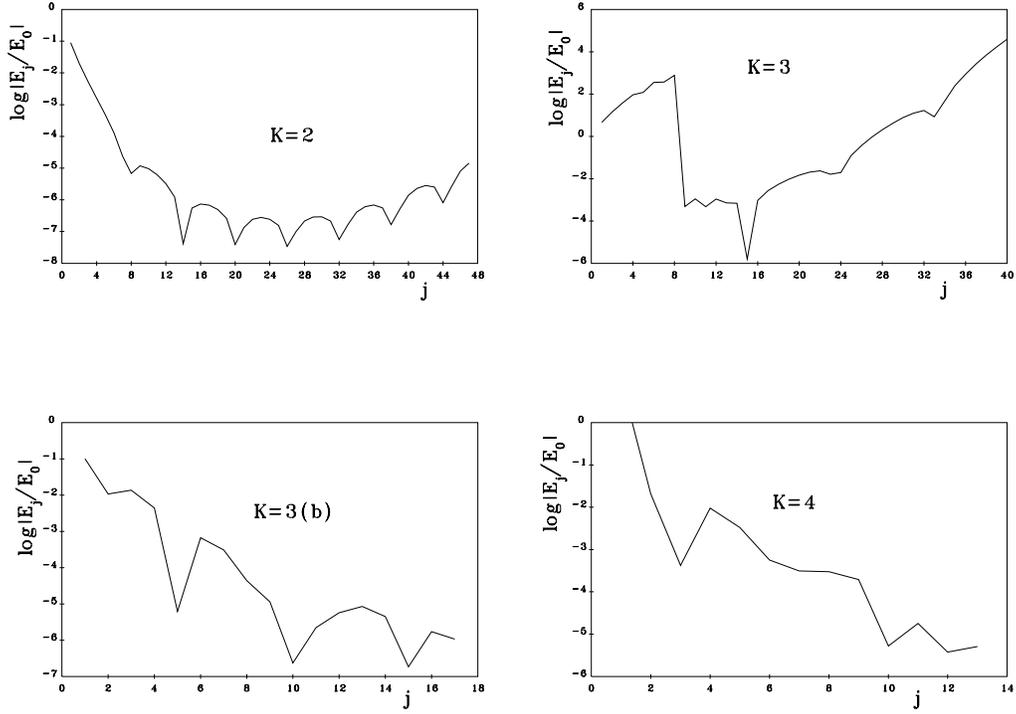}
\end{figure}

The systematic calculation of perturbation corrections of larger order offers
no difficulty if we resort to a computer algebra system like Maple
\cite{Maple}. Since we are unable to prove rigorously whether the perturbation
series converges, we resort to numerical investigation. Figure~\ref{fig:1}
shows that $\log |E_{j}/E_{0}|$ first decreases rapidly as $j$ increases, but
then it increases slowly suggesting that the series does not converge. If we
assume that the error on the energy estimated by the partial sum
\begin{equation}
E^{[M]}=\sum_{j=0}^{M}E_{j}  \label{eq:E[M]}
\end{equation}
is proportional to the first neglected term $|E-E^{[M]}|\approx |E_{M+1}|$,
then it is reasonable to truncate the perturbation series so that $|E_{M+1}|$
is as small as possible \cite{BO78}. In this case we find that $%
E_{26}=-0.3897686104\times 10^{-7}$ is the energy coefficient with the smallest
absolute value so that our best estimate is $E^{[25]}=1.06036215$ (compare with
the exact value in Table~\ref{tab:1}).

We proceed exactly in the same way for the pure sextic oscillator $K=3$.
Figure~\ref{fig:1} shows values of $\log |E_{j}/E_{0}|$ that clearly suggest
poorer convergence properties than in the preceding example. The energy
coefficient with the smallest absolute value is $E_{15}=0.2759118288\times
10^{-5}$, and our best estimate $E^{[14]}=1.14470$ is reasonably close to the
exact eigenvalue in Table~\ref{tab:1}. In principle, it is not surprising that
perturbation theory yields poorer results for $K=3$ than for $K=2$ \cite{F01}.

For the pure octic anharmonic oscillator $K=4$ we look for polynomial
solutions of the form
\begin{equation}
f_{j}(x)=\sum_{m=0}^{j+3}c_{j,2m+1}x^{2m+1},\;j=0,1,\ldots .
\label{eq:f_j_x8}
\end{equation}
In this case we have calculated less perturbation coefficients because they
require more computer memory and time. Surprisingly, the values of $\log
|E_{j}/E_{0}|$ in Figure~\ref{fig:1} suggest that the perturbation series for
$K=4$ exhibits better convergence properties than the one for $K=3$ just
discussed. The energy coefficient with the smallest absolute value (among those
we managed to calculate) is: $E_{12}=-0.5205493999\times 10^{-5}$ so that our
best estimate is $E^{[11]}=$ $1.225822$ which is quite close to the exact one
in Table~\ref{tab:1}.

The surprising fact that the convergence properties of the perturbation
series are clearly poorer for $K=3$ than for $K=4$ suggests that there
should be better solutions for the former case. If we try
\begin{equation}
f_{j}(x)=\sum_{m=0}^{j+2}c_{j,2m+1}x^{2m+1},\;j=0,1,\ldots
\label{eq:f_j_x6}
\end{equation}
then the values of $\log |E_{j}/E_{0}|$ are smaller than those obtained earlier
(compare $K=3\,(b)$ with $K=3$ in Figure~\ref{fig:1}). The coefficient with the
smallest absolute value is $E_{15}=-0.2344066313\times 10^{-6}$ and our best
estimate results to be $E^{[14]}=1.1448015$.

It is well known that Pad\'{e} approximants give considerably better results
than power series \cite{BO78}. We have tried diagonal Pad\'{e} approximants $%
[N,N]$ on the perturbation series for the cases $K=2$, $K=3\,(b)$, and $K=4$
and show results in Table~\ref{tab:1}. Notice that the Pad\'{e} approximants sum the $%
K=2$ series to a great accuracy but they are less efficient for $K=3\,(b)$
and $K=4$. This is exactly what is known to happen with the standard
perturbation series for anharmonic oscillators \cite{GG78}. However,
Pad\'{e} approximants appear to improve the accuracy of present perturbation
results in all the cases discussed above.

\section{\label{sec:Conclusions}Conclusions}

Present numerical investigation on the perturbation method proposed by
Bessis and Bessis \cite{BB97} suggests that although the series may be
divergent they are much more accurate than those derived from standard
Rayleigh--Schr\"{o}dinger perturbation theory. One obtains reasonable
eigenvalues for difficult anharmonic problems of the form $V(x)=x^{2K}$, and
results deteriorate much less dramatically than those from the standard
Rayleigh--Schr\"{o}dinger perturbation series as the anharmonicity exponent $%
K$ increases. In order to facilitate the calculation of perturbation
corrections of sufficiently large order we restricted our analysis to
polynomial solutions that are suitable for the ground state. The treatment
of rational solutions for excited states (like those considered by BB \cite
{BB97}) is straightfordward but increasingly more demanding.

Following BB \cite{BB97} we have implemented perturbation theory by
transformation of the linear Schr\"{o}dinger equation into a nonlinear
Riccati one. In this way the appropriate form of each potential coefficient $%
W_{j}$ reveals itself more clearly as shown in Section \ref{sec:Results} for
the quartic model. However, in principle one can resort to any convenient
algorithm because the perturbation method is sufficiently general as shown
in Section 2.

A remarkable advantage of the method of BB \cite{BB97}, which may not be so
clear in their paper, is its extraordinary flexibility as shown by the two
solutions obtained above for the case $K=3$. Moreover, the method of BB,
unlike two other renormalization approaches derived above as particular
cases \cite{K81,KGJ01}, does not require and adjustable parameter to give
acceptable results.

We believe that further investigation on the perturbation method of BB \cite
{BB97} will produce more unexpected surprises.

\begin{table}
\caption{Pad\'e approximants $[N,N]$ for the ground states of the quartic
($K=2$) sextic ($K=3 \,(b)$) and octic ($K=4$) oscillators} \label{tab:1}
\begin{tabular}{|c|c|c|c|} \hline
 $N$ &  $K=2$             & $K=3 \,(b)$            &  $K=4$       \\
 \hline
  1 &   1.06099633526211 &  1.13079274107556 &  1.210520271 \\
  2 &   1.06035870794451 &  1.14501385704172 &  1.203379654 \\
  3 &   1.06036222079274 &  1.14479406990471 &  1.225903753 \\
  4 &   1.06036204322721 &  1.14480776322951 &  1.225811686 \\
  5 &   1.06036210029130 &  1.14480229840545 &  1.225826780 \\
  6 &   1.06036209058862 &  1.14480243489611 &  1.225839331 \\
  7 &   1.06036209060731 &  1.14480245393688 &              \\
  8 &   1.06036209048882 &  1.14480245334362 &              \\
  9 &   1.06036209048295 &                   &              \\
 10 &   1.06036209048246 &                   &              \\
 11 &   1.06036209048301 &                   &              \\
 12 &   1.06036209048423 &                   &              \\
 13 &   1.06036209048420 &                   &              \\
 14 &   1.06036209048417 &                   &              \\
 15 &   1.06036209048417 &                   &              \\
 16 &   1.06036209048408 &                   &              \\
 17 &   1.06036209048418 &                   &              \\
 18 &   1.06036209048418 &                   &              \\
 19 &   1.06036209048418 &                   &              \\
 20 &   1.06036209048417 &                   &              \\
 21 &   1.06036209048418 &                   &              \\
 22 &   1.06036209048418 &                   &              \\
 23 &   1.06036209048418 &                   &              \\
 \hline
 Exact &1.060362090484183&1.14480245380      & 1.22582011\\
 \hline
\end{tabular}
\end{table}

\end{document}